# «Neutrino-4» experiment: preparations for search for sterile neutrino at 100 MW reactor SM-3 at 6-13 meters

A.P. Serebrov[a*], A.K. Fomin[a], V.G. Zinoviev[a], Yu.E. Loginov[a], M.S. Onegin[a], A.M. Gagarsky[a], G.A. Petrov[a], V.A. Solovey[a], A.V. Chernyi[a], O.M. Zherebtsov[a], V.P. Martemyanov[b], V.G. Zinoev[b], V.G. Tarasenkov[b], V.I. Alyoshin[b], A.L. Petelin[c], S.V. Pavlov[c], M.N. Svyatkin[c], A.L. Izhutov[c], S.A. Sazontov[c], D.K. Ryazanov[c], M.O. Gromov[c], N.S. Khramkov[c], V.I. Ryikalin[d]

[a]*PNPI, Gatchina, 188300 Russia*

[b]*NRC «Kurchatov Institute», Moscow, 123182 Russia*

[c]*JSC «SSC RIAR», Dimitrovgrad, 433510 Russia*

[d]*SRC IHEP, Protvino, 142281 Russia*

## Abstract

There has been designed an experimental project «Neutrino-4» for 100 MW reactor SM-3 to test the hypothesis of the «reactor antineutrino anomaly» [1, 2]. Advantages of the reactor SM-3 for such an experiment are low background conditions as well as small dimensions of a reactor core – 35x42x42 $cm^3$. One has carried on the Monte-Carlo modeling of a position sensitive antineutrino detector consisting of 5 operation sections, which as a result of displacement, covers the distance from 6 to 13 meters from the reactor core. One has succeeded in obtaining an experimental area of sensitivity to oscillation parameters $\Delta m^2$ and $\sin^2 2\theta$, which enables to verify the hypothesis of reactor antineutrino oscillations into a sterile state.

[*] E-mail: serebrov@pnpi.spb.ru

## Introduction

At present the possibility of existence of a sterile neutrino is being widely discussed. It is assumed that due to reactor antineutrino transitions into a sterile state one can observe both the effect of oscillations at short distances from the reactor and defict of reactor antineutrino flux at long distances [1, 2].

As early as 1967 B.M. Pontekorvo initially assumed [3,4] that neutrino transitions into a sterile state were likely to occur. Further development of the idea of neutron oscillations allowed describing neutrino oscillations with the model of three neutron generations. The picture of this phenomenon can be sufficiently well represented by the matrix of Pontecorvo-Maki-Nakagawa-Sakata [5]. However there is available a number of experimental facts which point out necessity of extending this scheme.

The first of them is related to the so-called LSND anomaly [6], which was later on investigated in MiniBooNE and MINOS [7,8] experiments. Another experimental evidence is Ga anomaly [9,10], that emerged in calibrating Ga neutrino detectors. Some indication of necessity for introducing an additional neutrino type follows from the analysis of initial nuclear synthesis processes [11] as well as from large scale structures being formed in the Universe [12,13]. In addition, sterile neutrinos are regarded as being candidates for dark matter in the Universe [13].

Finally in the early 2011 a reactor antineutrino anomaly was claimed [1,2]. An additional analysis of data relevant to neutrino production in reactors has shown that the calculated neutrino flux should be increased by 3%. Thus deficiency of registered events in neutrino experiments has emerged. Some contribution (~0.7%) into this deficiency has been made by alteration of neutron lifetime in correspondence with new experimental data [14]. Neutron lifetime has decreased approximately by 1% [15], correspondingly, efficiency of neutrino detectors has increased by 1%, since efficiency of neutron detectors is formed by cross-section of the reaction of reverse neutron beta-decay. As a result, ratio of the experimentally observed neutron flux to the predicted one has changed from 0.976±0.024 to 0.943±0.023 [1]. The effect concerned makes up 2.5 of standard deviation. This is not yet sufficient to be sure of existence of reactor antineutrino anomaly. It should be noted that the effects discussed earlier are also at the validity level 2.5 – 3.0 of standard deviations, therefore performance of new and more accurate experiments are considered to be extremely significant.

## SM-3 reactor

We have studied opportunity of conducting new experiments using research reactors in Russia. The most favorable conditions for carrying out an experiment on search of neutrino oscillations at short distances can be provided by the reactor SM-3 (Fig. 1). Advantages of the reactor SM-3 are low background conditions, the reactor compact zone – 35x42x42 $cm^3$ at high reactor intensity – 100 MW, as well as quite a small distance from the center of the reactor core to the experimental hall wall – 5 m. Besides, of great significance is the fact that neutrino flux can be measured within sufficiently wide distance range – from 6 to 13 meters. One can expect up to $3{\cdot}10^3$ neutrino events per day to occur at the reactor intensity 100 MW at the distance of 6 m from the reactor core within the volume 1 $m^3$.

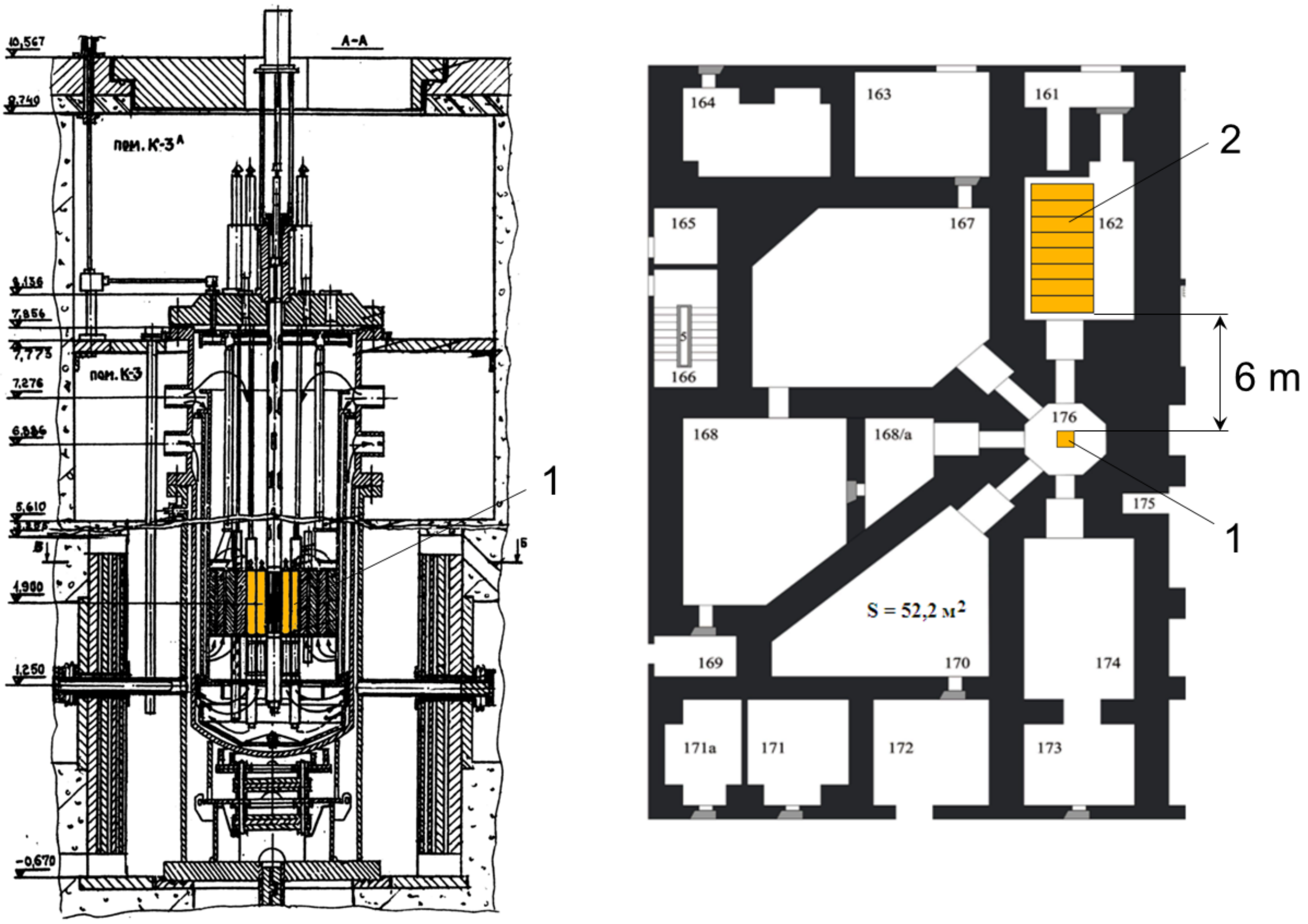

Fig. 1. The experimental outlay at the SM-3 reactor: 1 – reactor core, 2 – antineutrino detector.

Initially 100-MW reactor SM-3 was designed for conducting both beam and loop experiments. There were constructed 5 beam buildings separated from each other by solid concrete walls of ~1 m width. It enabled to carry out experiments on neutron beams without changing background conditions at the neighboring installations. Further the main experimental program was focused on the tasks connected with radiation in the reactor core. 25 years later rather high fluence was accumulated on the reactor case materials which caused the necessity of its replacing. The simplest solution was to install a new reactor case on the old reactor tank. However, according to this solution the reactor core had to be risen 67 cm higher than its former location. Horizontal beam channels had to be removed because of priority of loop and radiation experiments. Neutron flux at the location of the former beam channels was decreased by four orders of value. Correspondingly, neutron background in the former beam halls was diminished. It comprised approximately $10^{-3}$ n/($cm^2$s) (thermal neutrons). This is nearly 4-5 order of value less than the typical neutron background in the beam hall of the research reactor. Recently with the aim of preparing an experiment on search for transitions of reactor antineutrino into sterile state at the reactor SM-3 the gate installation of the former neutron beam was upgraded. It resulted in the background of rapid neutrons being decreased up to several units of $10^{-4}$ n/(cm$^2$s) (fast neutrons), i.e. practically up to neutron fluxes on the Earth surface induced by outer space radiation.

## Simulation of the reactor antineutrino detector

Applying the Monte Carlo method we have simulated a reactor antineutrino detector. The detector of scintillation type is based on using reaction $\tilde{\nu}_e + p \rightarrow e^+ + n$. At first the detector registers a positron whose energy is determined by that of antineutrino and 2 annihilation gamma quants, each having energy 511 keV. Neutrons emerging in the reaction are absorbed by Gd to form a cascade of gamma quants, with total energy being about 8 MeV. The detector will register two subsequent signals produced by positron and neutron. Mineral oil ($CH_2$) with addition of Gd 1 g/l is used as a scintillator material. The scintillator light emission is equal to $10^4$ photons for 1 MeV. The detector is multi-sectoring for measuring dependence of antineutrino flux and spectrum at the reactor core distance. In case of detecting distance effect of oscillations one should make analysis of neutrino flux spectral alterations depending on distance. Only self-consistent effects in space and spectrum can be regarded as evidence for existence of transitions into a sterile state. The antineutrino spectrum will recover from the positron spectrum, since in the first approximation the relationship between positron energy and that of antineutrino is linear: $E_{\tilde{\nu}} = E_{e^+} + 1.8\,\mathrm{MeV}$.

The detector outlay is shown in Fig. 2. The detector consists of 5 sections 1x0.8x0.4 m$^3$ with partitions rigidly fixed between them. Partitions are applied to prevent light from emitting outside sections. Photomultipliers PMT-49B located on the upper surface are employed in the detector. Between PMT and a scintillator there is an air layer for leveling down light collection. It reduces light collection but improves energy resolution of detector.

The problem of various efficiency in detecting separate sections can be solved by displacing the detector. It will allow to do relative calibration of sections among themselves. Fig. 3 shows a possible difference in efficiency of sections and obtaining one dependence after calibration of efficiency of sections among themselves.

In simulation the antineutrino beam was supposed to direct parallel to the detector axis. The direction of positron emission was considered to be isotropic. Space distribution of neutron capture by Gd, as well as calculaions of energy release from positron and gamma quants in the scintillator were made with program MCNP [16]. The spectrum of gamma quants in neutron capture by Gd was generated for reaction $^{157}$Gd(n,$\gamma$). Exponential time of flight of photons in the scintillator is 4 m. Photons are reflected from the walls with probability 0.95.

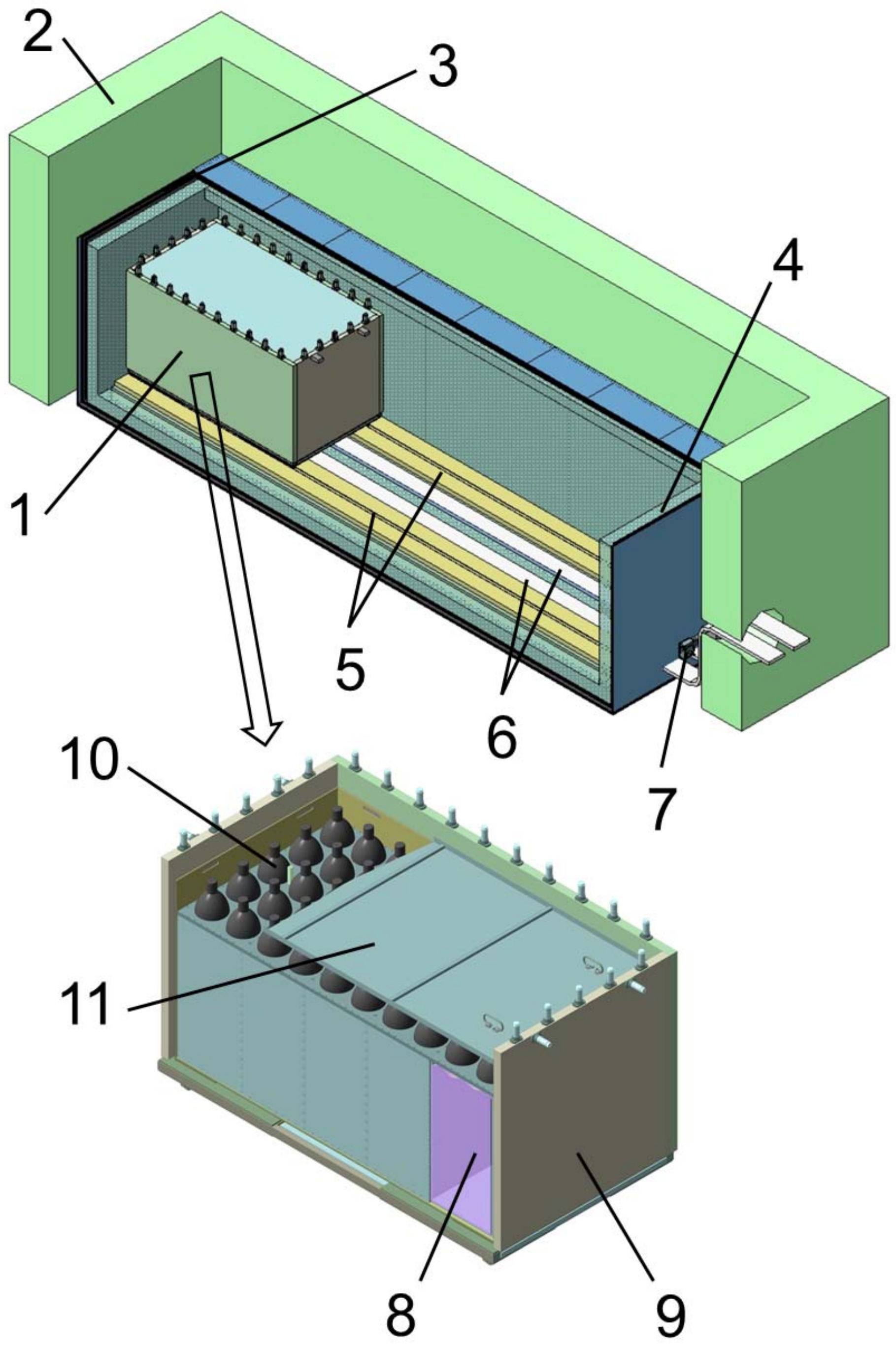

Fig. 2. The detector outlay: 1 – detector of the reactor antineutrino, 2 – building walls, 3 – lead shielding 6 cm, 4 – shielding made of boron polyethylene 16 cm, 5 – rails, 6 – cable channels, 7 – motor for displacing a detector, 8 – separate detector section with scintillation liquid, 9 – scintillation plates of internal anticoincidence shielding with PMT, 10 – detector PMT, 11 – tank cover.

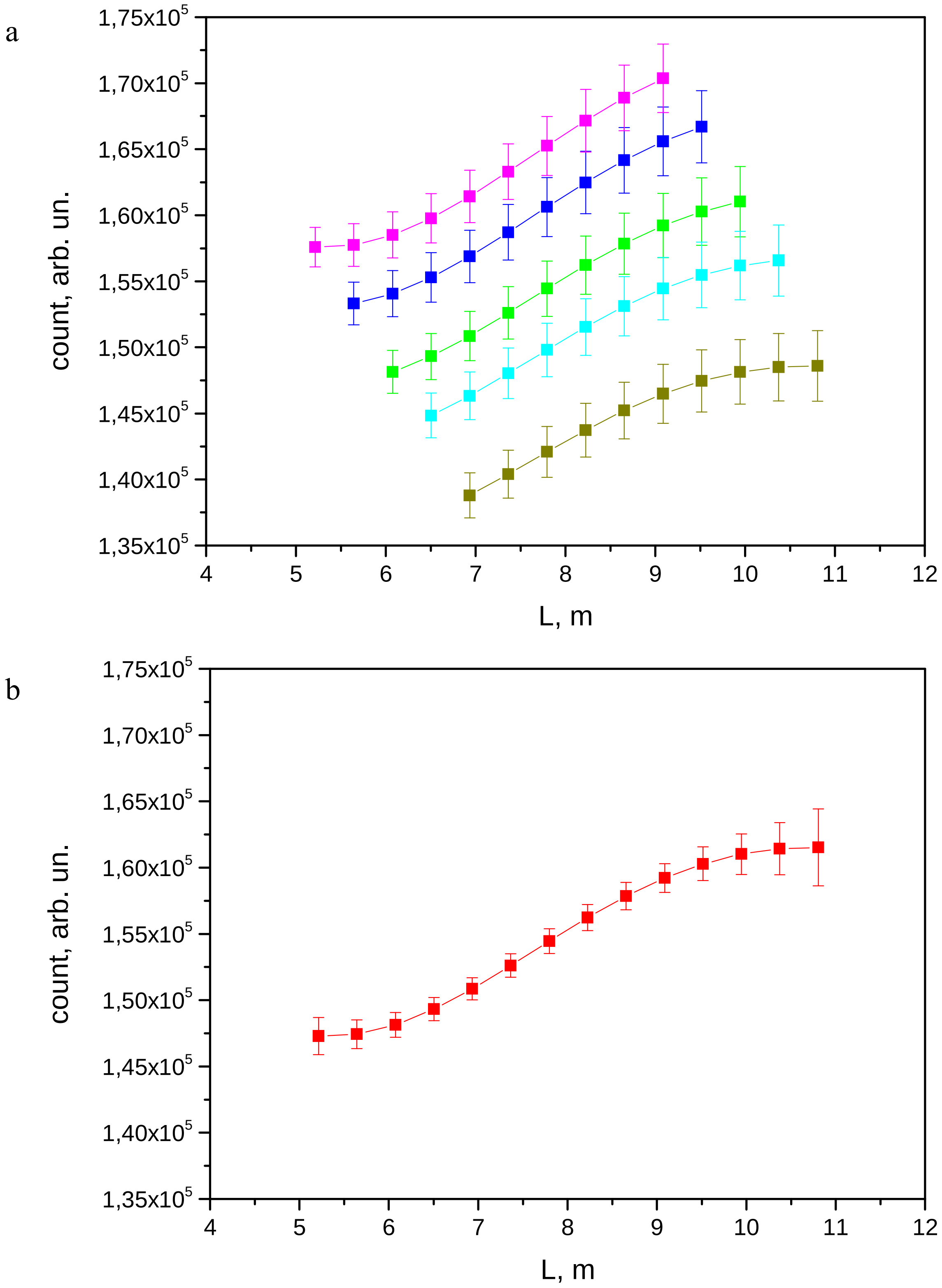

Fig. 3. Sample of oscillation curve measurement for 1 year with $\Delta m^2 = 1$ eV$^2$ and $\sin^2 2\theta = 0.15$: a – curves obtained by each of 5 detector sections, b – one dependence after calibration of efficiency of sections among themselves.

Fig. 4. shows distribution of PMT signal (the number of registered photons) from positrons of different energy. Respectively, one can restore positron energy by counting the number of registered photons using relationship between the count of photons and energy presented in Fig. 5. Fig. 6. shows the number of registered photons depending on the place of positron birth.

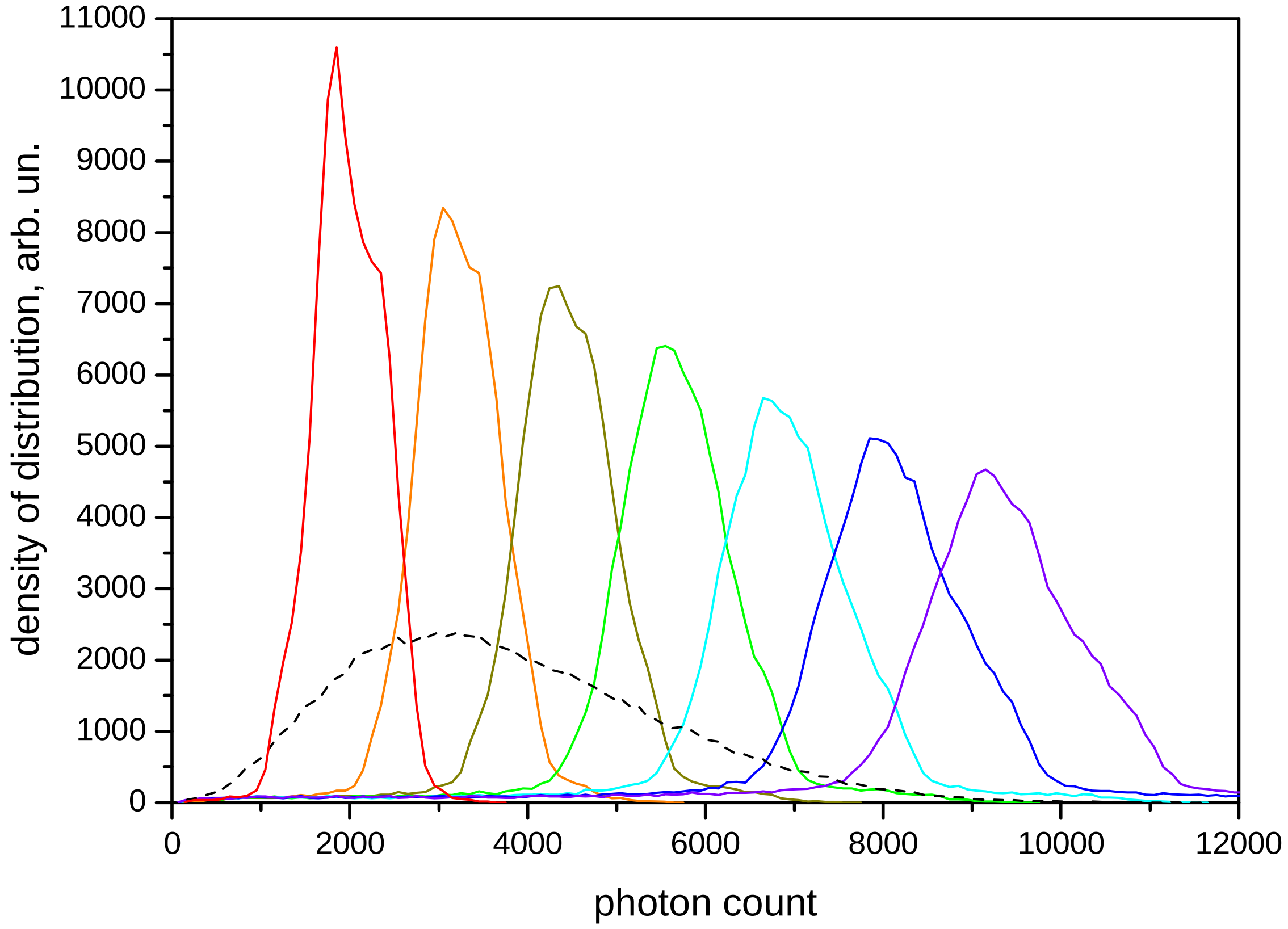

Fig. 4. PMT count distributions from positrons with energy from 1 to 7 MeV. Shaded line is the distribution for positron spectrum.

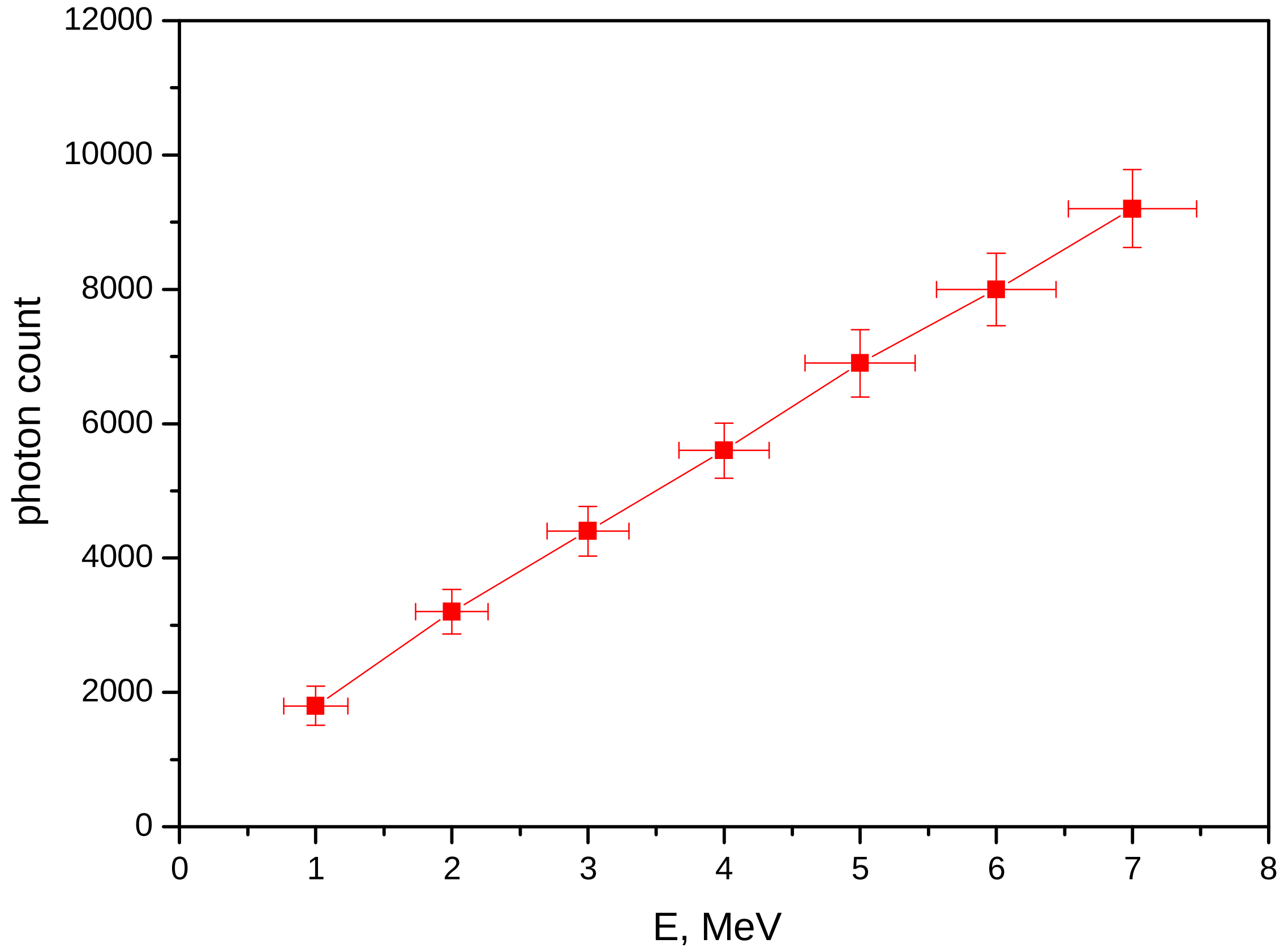

Fig. 5. Dependence for restoring energy of a positron on PMT count.

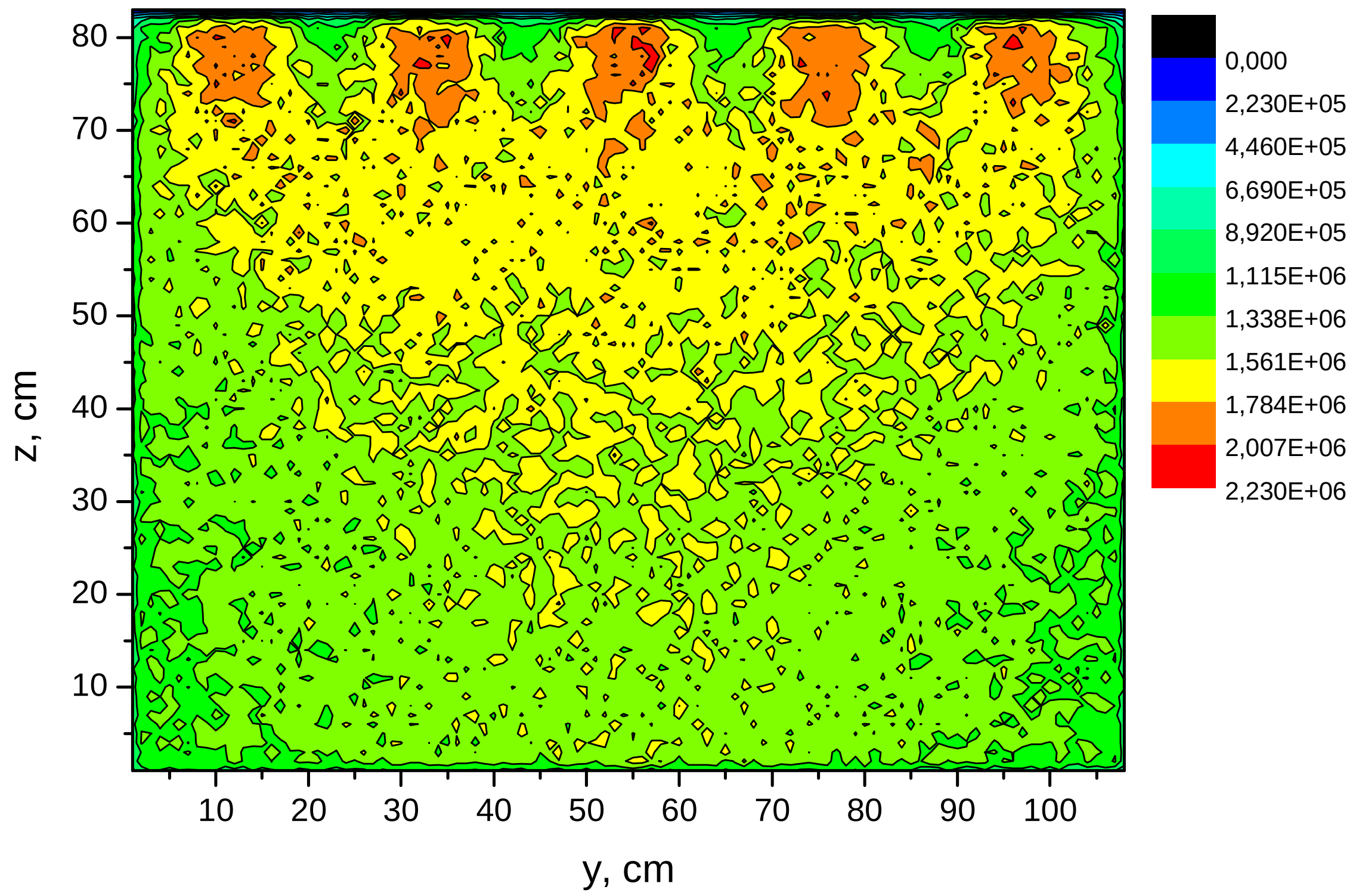

Fig. 6. Count of PMT depending on the place of positron birth with energy 4 MeV.

Simulation resulted in obtaining distribution of the number of photons that attained all PMT (light collection) from positrons at different energies on account of energy release by gamma quants. Light collection for the section is ~12 %. Electron signal with PMT will be proportional to the number of photons reaching the photo cathode, thus it is possible to recover approximately positron energy. Although the number of photoelectrons is noticeably less than that of photons, their statistics is still high enough to disregard dispersion of photo electron emission. Later on we will make analysis using units of the number of photons that reached all PMT of one section. In order to suppress background related to the neutron capture by hydrogen it is reasonable to choose the positron signal threshold of 2700 photons (~2.2 MeV). Efficiency of detecting positron events in such conditions depends on positron energy (Fig. 7). For the positron spectrum it comprises $\varepsilon_{e^+}$ =0.688(5). The section where antineutrino registration is made will be determined according to the positron event as it will have more localised energy release as compared with the neutron one.

In detecting a neutron signal the threshold of 6150 photons (~5 MeV) is assumed to be used, which enables to supress the background of random coincidences, with natural radioactivity not exceeding 4 MeV [17]. It is essential to consider the total count PMT of all detector sections in the coincidence mode. Detection efficiency of a neutron signal in the middle detector section is equal to $\varepsilon_n$=0.616(5). In view of ~20% neutrons being captured by hydrogen with emission of about 2.2 MeV [4], $\varepsilon_n$=0.493(5). The detector efficiency in the middle section obtained as a result of modeling is equal to $\varepsilon$=0.348(5). Fig. 8 shows the expected distributions of positron and neutron signals.

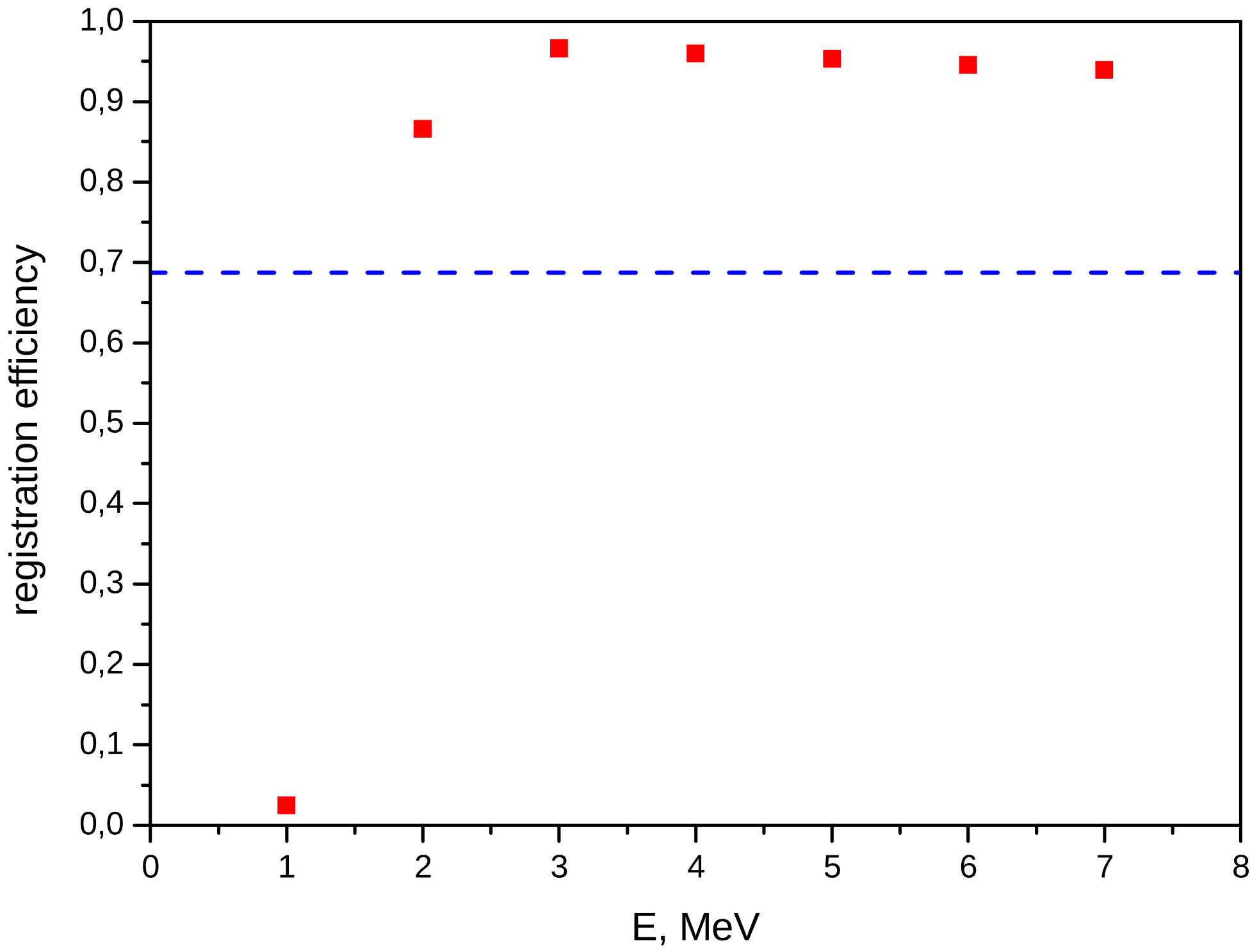

Fig. 7. Efficiency of positron event registration depending on the positron energy. Dotted line is registration efficiency for the whole spectrum of positrons.

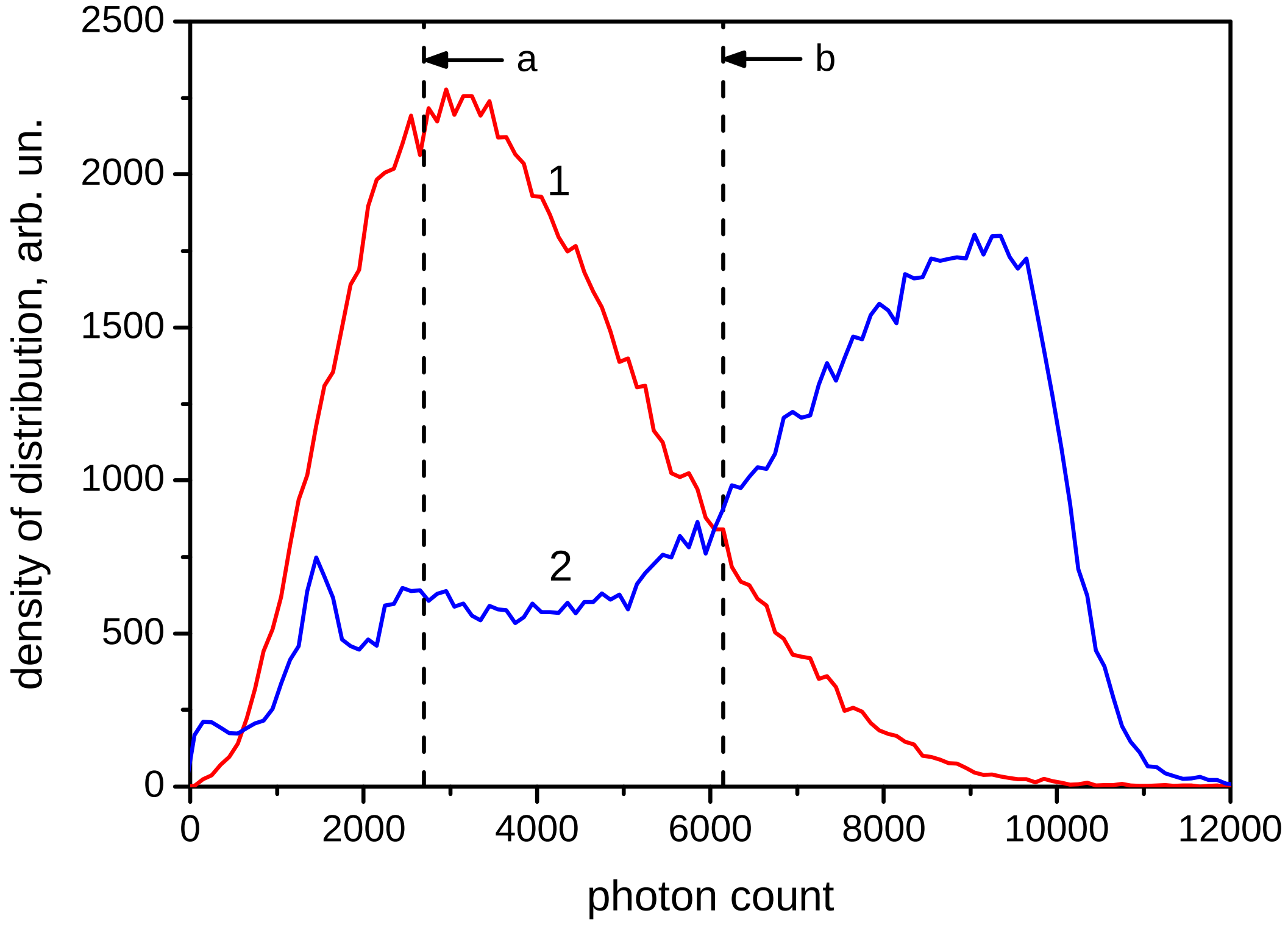

Fig. 8. Expected distribution of signals from positrons (1) and neutrons (2). a – threshold of positron signal 2700 photons (~2.2 MeV); b – threshold of neutron signals 6150 photons (~5 MeV).

Uniform distribution of antineutrino events along the detector length has been modeled. In lateral sections detection efficiency is lower due to possible escaping of positron and neutron outside the detector through the lateral surface and also due to the fact that part of neutron events is lost at the threshold 5 MeV because of gamma quants escaping outside the detector through the lateral surface (Fig. 9).

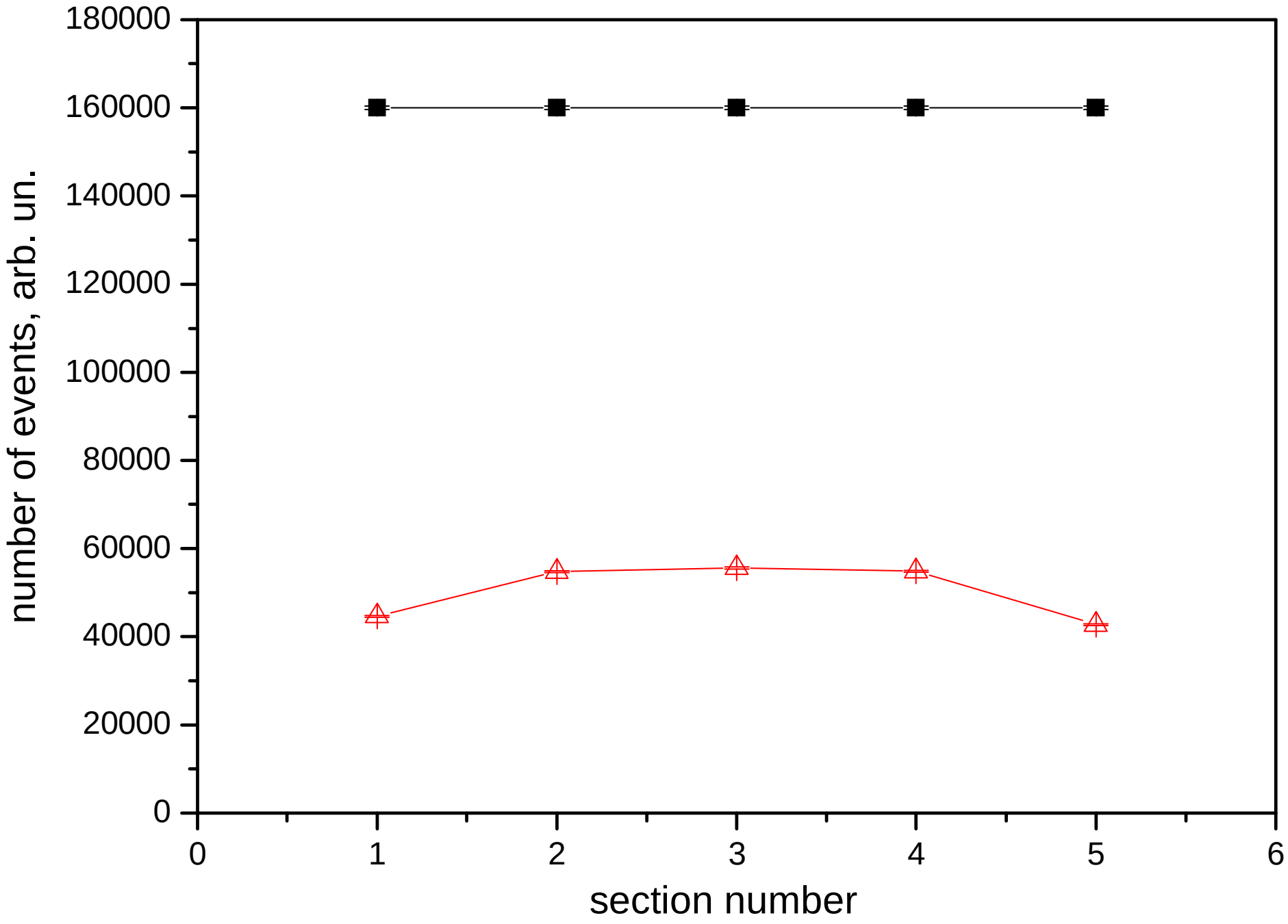

Fig. 9. Distribution of events among sections. ■ is actual events (uniform distribution), Δ is determination of the event location as a result of modeling.

**Sensitivity of «Neutrino-4»**

The oscillation process is described by the equation $P(\tilde{\nu}_e \rightarrow \tilde{\nu}_e) = 1 - \sin^2 2\theta \ \sin^2(1.27 \frac{\Delta m^2 [\mathrm{eV}^2] L[\mathrm{m}]}{E_{\tilde{\nu}}[\mathrm{MeV}]})$ [1]. To detect an experimental sensitivity $\chi^2$ is calculated for the hypothesis according to which oscillations are missing. Sensitivity of experiments on search of neutrino oscillations can be determined by integral and differential methods. In terms of an integral method the measured antineutrino flux and spectrum are compared with the calculated ones. In terms of a differential method the measured antineutrino flux and spectrum are compared with those between different detector sections. In determining the area of experimental sensitivity we applied a differential method, since in integral method there are uncertainties concerned with information on the calculated antineutrino spectrum and flux and that on detector efficiency. Fig. 10 shows the sensitivity area of the experiment during a year of collecting statistics at 95% CL obtained by a differential method. The area in question covers the most probable values of oscillation parameters $\Delta m^2$ and $\sin^2 2\theta$ [1].

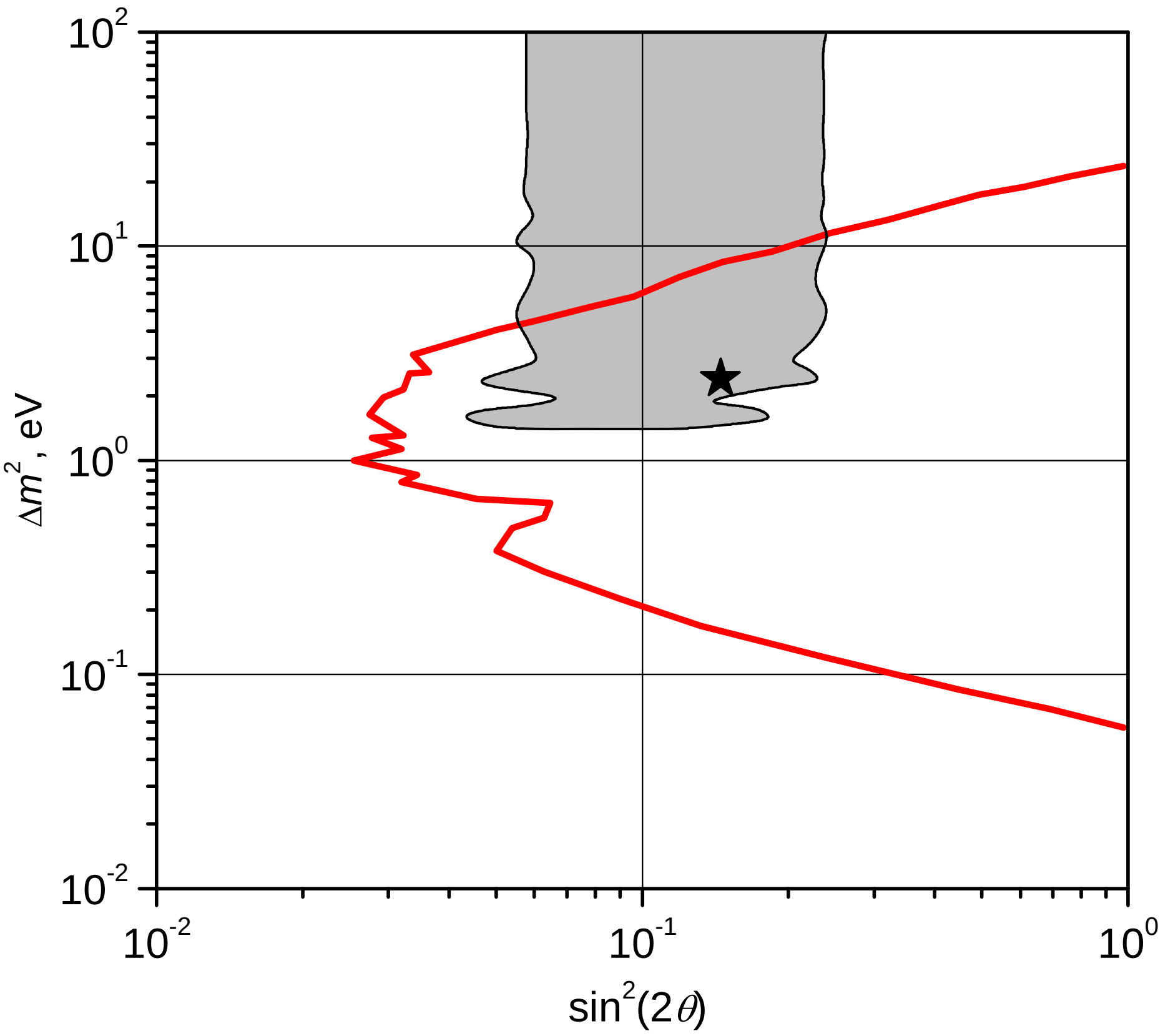

Fig. 10. Solid line - experimental sensitivity area of «Neutrino-4» for oscillation parameters during a year of collecting statistics at 95% CL. Shaded area – constraints on the oscillation parameters $\Delta m^2$ and $\sin^2 2\theta$ at 95% CL [1], ★ – the most probable parameter values.

**Estimation of the effect–background relation from the experiment [17]**

The experiment made by a research team under L.A. Mikaelyan [17] was devoted to investigation of reverse beta decay reaction $\tilde{\nu}_e + p \to e^+ + n$. It was carried on at the Rovenskaya atomic station (1375 MW) at the distance of 18 m from the reactor core. The antineutrino flux in the detector location was equal to $6{\cdot}10^{12}$ cm$^{-2}$s$^{-1}$, which is only 1.5 higher than that in the first section of our detector. The volume of a liquid scintillator with Gd addition was 238 l, i.e. by ~30% less than that of the designed detector section. The experiment was carried out on the surface of the Earth, though it should be noted that the whole atomic station construction served as a passive shielding from outer space radiation. On the whole the conditions of the performed experiment and that to be performed are rather similar, therefore we are able to take into account the results of the conducted experiment [17]. We are also planning to use both an external shielding (relative to passive shielding of the detector) and an internal active shielding from outer space radiation. Fig. 2 shows only an internal active shielding. External active shielding will be located over the detector and will consist of 60 units.

The background of random coincidences in the experiment [17] was greatly dependent on amplitude of recorded signals. The ratio of effect to the background of random coincidences was equal to 1 at energy ~2 MeV. Below this boundary the background of random coincidences sharply increased but at energy ~3 MeV it dropped much lower the neutrino effect. Thus the threshold 5 MeV for a neutron signal was chosen with sufficiently high excess in simulation our detector. One can expect statistics of neutrino effects to play a determining role at the threshold 2.2 MeV for a positron signal and such is the case for a neutron signal at the threshold 5 MeV.

## Conclusion

There has been designed an experimental project «Neutrino-4» for 100 MW reactor SM-3 to search reactor antineutrino oscillations into a sterile state. MC model of reactor antineutrino detector is built. The detector efficiency in the middle section obtained as a result of modeling is equal to $\varepsilon$=0.348(5). Thus efficiency of detecting positron events is $\varepsilon_{e^+}$=0.688(5) (threshold ~2.2 MeV), efficiency of detecting neutron events is $\varepsilon_n$=0.493(5) (threshold ~5 MeV). The sensitivity area of the experiment during a year of collecting statistics at 95% CL is obtained. The area in question covers the most probable values of oscillation parameters $\Delta m^2$ and $\sin^2 2\theta$.

This work has been carried out at the support of the Ministry of education and science of the Russian Federation, agreement № 8702. The investigation has been supported by Russian Foundation for Basic Research, grant № 12-02-12111-ofi_m.

## References

[1] G. Mention, M. Fechner, Th. Lasserre et al., Phys. Rev. D 83, 073006 (2011).

[2] T. Mueller, D. Lhuillier, M. Fallot et al., Phys. Rev. C 83, 054615 (2011).

[3] B. Pontecorvo, Zh. Exp. Teor. Fiz. 33, 549 (1957); 34, 247 (1958).

[4] B. Pontecorvo, Zh. Exp. Teor. Fiz. 53, 1717 (1967).

[5] Z. Maki, M. Nakagava, S. Sakata, Prog. Theor. Phys. 28, 870 (1962).

[6] A. Aguilar et al. (LSND coll.), Phys. Rev. D 64, 112007 (2001).

[7] A.A. Aguilar-Arevalo et al. (MiniBooNE coll.), Phys. Rev. Lett 103, 111801 (2009); 105, 181801 (2010).

[8] P. Adamson et al. (MINOS coll.), Phys. Rev. Lett. 107, 021801 (2011).

[9] V.N. Gavrin, V.V. Gorbachev, E.P. Veretenkin et al., arXiv:1006.2103.

[10] C. Giunti, M. Laveder, Phys. Rev. C 83, 065504 (2011).

[11] Y. Izotov, T. Thuan, Astrophys. J. Lett. 710, L67 (2010).

[12] J. Hamann, S. Hannestad, G.G. Raffelt et al., arXiv:1108.4136.

[13] A. Kusenko, Phys. Rept. 481, 1 (2009).

[14] A.P. Serebrov, V.E. Varlamov, A.G. Kharitonov et al., Phys. Lett. B 605, 72 (2005);
A.P. Serebrov, V.E. Varlamov, A.G. Kharitonov et al., Phys. Rev. C 78, 035505 (2008).

[15] A.P. Serebrov, A.K. Fomin, Phys. Rev. C 82, 035501 (2010).

[16] Briesmeister J. F., Ed., "MCNP - A General Monte Carlo N-Particle Transport Code, Version 4C," LA-13709-M (April 2000).
[17] Afonin A.I., Ketov S.N., Kopeikin V.I. et al., Zh. Eksp. Teor. Fiz. 94, 1 (1988);
Klimov Yu.V., Kopeikin V.I., Labzov A.A. et al., Yad. Fiz. 51, 401 (1990).